\documentclass[12pt]{article}
\usepackage{makeidx}
\usepackage{amssymb}
\usepackage{amsfonts}
\usepackage{amsmath}
\usepackage{graphicx}
\usepackage{setspace}
\usepackage{authblk}
\usepackage{units}
\usepackage{appendix}
\usepackage[left=2cm,top=2.3cm,right=2cm,bottom=2.1cm]{geometry}
\usepackage{xcolor}
\usepackage[hyperfootnotes=false]{hyperref}

\renewcommand\familydefault\rmdefault
\newcommand{\lambdabar}{{\mkern0.75mu\mathchar '26\mkern -9.75mu\lambda}}
\hypersetup{colorlinks=true,linkcolor=black,citecolor=black,urlcolor=black}

\begin{document}

\title{{\textbf{Saturation of Energy Levels of the Hydrogen Atom in Strong Magnetic Field}}}
\author[1,2,3]{T. C. Adorno\thanks{adorno@hbu.edu.cn, tg.adorno@gmail.com}}
\author[3,4,5]{D. M. Gitman\thanks{gitman@if.usp.br}}
\author[3,4]{A. E. Shabad\thanks{shabad@lpi.ru}}
\affil[1]{\textit{Department of Physics, College of Physical Sciences and Technology, Hebei University, Wusidong Road 180, 071002, Baoding, China;}}
\affil[2] {\textit{Key Laboratory of High-precision Computation and Application of Quantum Field Theory of Hebei Province, Hebei University, Baoding, China;}}
\affil[3]{\textit{Department of Physics, Tomsk State University, Lenin Prospekt 36, 634050, Tomsk, Russia;}}
\affil[4]{\textit{P. N. Lebedev Physical Institute, 53 Leninskiy prospekt,
119991, Moscow, Russia;}}
\affil[5]{\textit{Instituto de F\'{\i}sica, Universidade de S\~{a}o Paulo, Caixa Postal 66318, CEP 05508-090, S\~{a}o Paulo, S.P., Brazil.}}

\maketitle

\onehalfspacing 
\begin{abstract}
We demonstrate that the finiteness of the limiting values of the lower energy
levels of a hydrogen atom under an unrestricted growth of the magnetic
field, into which this atom is embedded, is achieved already when the vacuum
polarization (VP) is calculated in the magnetic field within the
approximation of the local action of Euler--Heisenberg. We find that the
mechanism for this saturation is different from the one acting, when VP is
calculated via the Feynman diagram in the Furry picture. We study the
effective potential that appears when the adiabatic (diagonal) approximation
is exploited for solving the Schr\"{o}dinger equation for the longitudinal
degree of freedom of the electron on the lowest Landau level in the atom. We find that the (effective) potential of a point-like charge remains
nonsingular thanks to the growing screening provided by VP. The~regularizing
length turns out to be $\sqrt{\alpha /3\pi }\lambdabar _{\mathrm{C}}$, where $\lambdabar _{\mathrm{C}}$ is the electron Compton length. The~family of effective potentials, labeled by growing values of the magnetic field
condenses towards a certain limiting, magnetic-field-independent
potential-distance curve. The~limiting values of even ground-state energies
are determined for four magnetic quantum numbers using the Karnakov--Popov method.
\end{abstract}

%%%%%%%%%%%%%%%%%%%%%%%%%%%%%%%%%%%%%%%

\section{Introduction}
It is known since the papers by V. V. Usov and one of the present authors 
\cite{ShaUso07,ShaUso071} that the account of radiative corrections cures the unwanted
property of the ground state of a hydrogen-like atom---peculiar to quantum
mechanics (QM)---to become infinitely deep when an external constant
magnetic field, to which this atom is exposed, tends to infinity.

The~reason for this property, first established long ago by R.J. Loudon \cite%
{EllLou60,EllLou6066}, lies in the fact that the electron degree of freedom transverse
to the magnetic field becomes extinct, since in the transverse plane, the
electron is restricted to the lowest Landau orbit. As~for the remaining
londitudinal degree of freedom, it is governed---when the magnetic field is
sufficiently large to dominate over the Coulomb field and thereby to justify
the so-called adiabatic approximation~\cite{SnySch39}---by the
one-dimensional Schr\"{o}dinger equation in the Coulomb field of atom's
nucleus. In~the one-dimensional problem, the~Coulomb attraction proves
to be too singular and causes the \textquotedblleft fall down onto the
center\textquotedblright\ in the terminology of Ref.~\cite{Landauvol3}. As~a
result, the~ground level goes to negative infinity as logarithm squared of
the magnetic field.

The~matters change after the radiative correction, in~the form of the vacuum
polarization, is taken into consideration. The~point is that after the
polarization tensor had been \\ \mbox{calculated~\cite{BatSha71,BatSha712,BatSha713,BatSha714,BatSha715}} (\mbox{see also \cite%
{Shabad88}}) beyond the photon mass shell in the furry picture as a Feynman
loop of exact Dirac propagators of electron and positron in a constant
magnetic field, it was found~\cite{Shabad88,melrose} that its component
responsible for the longitudinal electrostatic screening grows linearly with
magnetic field\footnote{%
The~same asymptotic behavior was obtained independently in {Ref.} 
 \cite{skobelev,skobelev2}, by calculating the vacuum polarization (VP) diagrams in a special
two-dimensional electrodynamics conjectured to be presenting the
large-magnetic-field regime~of~QED.}. The~scalar potential produced by the
point-like electric charge was studied in a magnetic field using the photon
propagator obtained by summation of the one-photon-reducible chain of the
above-described VP diagrams in~\cite{ShaUso07,ShaUso071} and in \cite%
{sadooghi}. It was found that at large distances from the charge this
potential is nothing but the Coulomb law modified by the presence of
anisotropic medium---to which the magnetized vacuum is equivalent, in
accordance with the early statements~\cite{erber,Birula70}. However, close to the
charge, the~Coulomb singularity in the longitudinal part of the potential $%
1/x_{\parallel }$ is replaced (in the limit of infinite magnetic field) by
the delta function $\delta (x_{\parallel })$. This~sort of singularity does
not cause the \textquotedblleft fall down onto the center\textquotedblright
. Hence, the~limiting ground-state energy is finite~\cite{ShaUso07,ShaUso071}. The
physical reason for this effect is, certainly, the~growth of the
polarization with the field, which strongly screens the Coulomb field.

For finding the resulting finite value of the ground state, Machet and
Vysotsky~\cite{MacVys11} proposed an analytic interpolation for the
asymptotic form of the polarization operator that allowed them to reproduce
computer-calculated potential curves of Ref.~\cite{ShaUso07,ShaUso071}. This
interpolation also allowed them to efficiently use the method of
Karnakov--Popov~\cite{PopKar03} (KP) to correct a rough preliminary estimate
for the limiting value of the ground state energy made in~\cite{ShaUso07,ShaUso071}.
Later, this~method was further elaborated to include the study of excited
states~\cite{PopKar12,PopKar14}. In~\cite{godunov}, the~critical
charge dependence\footnote{By critical charge of a nucleus, its value is meant when the ground state of a hydrogen-like atom treated via the Dirac equation sinks into the lower continuum, see for instance~\cite{Greiner}.} on the magnetic field found initially in~\cite{semikoz} was studied as influenced by the strong screening phenomenon under discussion for extended charge (a nucleus of final size) using KP method. We shall also be using KP procedure in the present paper.

The~present work is based on the observation that, in~practice, the
finiteness of the ground state energy may be obtained\ already based on the
local effective action functional. In~the one-loop approximation, it is
known as the action of Euler--Heisenberg (EH)~\cite{HeiEul36}. The
polarization tensor obtained by variational differentiation of it over its
field arguments is valid as applied to sufficiently slowly varying fields.
This~restriction, of which the approach in~\cite{ShaUso07,ShaUso071,sadooghi,MacVys11,PopKar12,PopKar14} is free, is owing to the
fact that time- and space-derivatives of the fields are meant to be
disregarded in the course of calculation of the local action functional. The
large-magnetic-field asymptotic behavior of the EH action considered in \cite%
{heyl} gives the contribution to the \textquotedblleft
steady\textquotedblright\ polarization tensor, coinciding in the
small-momentum domain with the asymptotic form of the \textquotedblleft
genuine\textquotedblright\ polarization tensor. Correspondingly, the
large-distance Coulomb-like form of the potential obtained in~\cite{ShaUso07,ShaUso071,sadooghi} is readily reproduced~\cite{AdoGitSha16} within the EH
approach. For our present goals, it is essential that the linear growth with
the magnetic field of a particular component of polarization tensor retains
in this approach\footnote{%
Moreover, it can be demonstrated that this property is preserved by the
polarization tensor obtained from the two-loop expression for\ the local
effective action, as~calculated in~\cite{Ritus2}.}. Correspondingly, there
remains the large screening of the Coulomb field, and~it is enough for the
full freezing of the ground level. However, its mechanism is different from
the one that acts beyond the local action approach. In~the next Subsection,
we shall discuss this mechanism qualitatively. Our~conclusion will be that
due to the strong screening, the~electron wave function in the field of a
point charge remains concentrated at a finite distance from the charge,
hence the singularity is not achieved even when the magnetic field is
formally infinite. This~situation is opposed to the QM case of absence of
strong screening when the domain of electron localization, the~magnetic
length, shrinks to zero in that limit. In~that Subsection, we~also establish
the upper limits on the magnetic field to which our results are applicable.

In~Section \ref{Sec2}, we~consider the one-dimensional effective potential,
to which the anisotropic Coulomb law gives rise, and~which is involved in
the one-dimensional Schr\"{o}dinger equation occurring in the adiabatic
approximation of~\cite{SnySch39}. It is illustrated in Figure~\ref{F0} %MDPI: Wrong order of the figures, figure 2 and 4 were before figure 1, please check. [REPLY: The first figure (labelled as F0) is correctly cited now.]
 how the
potental-distance curves condense towards a certain limiting curve for large
magnetic fields instead of dropping down infinitely, unlike is the case
in QM. The~form of the limiting curve predetermines the finiteness of the
lower energy levels (as a matter of fact, of all excited levels,
as well), which is found in Section \ref{Sec2} using the KP method.
Numerical results and the resulting saturation pattern, shown in Figure~4 for the first even-energy levels, are discussed in\ Section \ref{Sec3}.%MDPI:Figure should be cited in numerical order, please check it [REPLY: It is impossible to shift figure 4 (labelled as F3) upwards to put it the correct numerical order (I mean, before figures 2 (labelled as F1) and 3 (labelled as F2)) because it belongs to the last section of the paper and we need to cite this particular Figure here. So, in the published version, it must appears Figure 4 here.]

In~this work, we~consider the four-dimensional Minkowski space-time,
parameterized by coordinates $x^{\mu }=\left( x^{0},\mathbf{r}=x^{i}\,,\
i=1,2,3\right) $ and metric tensor $\eta _{\mu \nu }=\mathrm{diag}\left(
+1,-1,-1,-1\right) $. The~electromagnetic field strength tensor $F_{\mu \nu
}=\partial _{\mu }A_{\nu }-\partial _{\nu }A_{\mu }$ and its dual $\tilde{F}%
^{\mu \nu }=\left( 1/2\right) \varepsilon ^{\mu \nu \rho \sigma }F_{\rho
\sigma }$ are related through the Levi--Civita antisymmetric tensor,
normalized as $\varepsilon ^{0123}=1$. The~Gaussian system of units, in
which $\alpha =e^{2}/\hslash c$, is employed throughout the text.

\section{Spectrum Equation for Lower Even-Energy Levels\label{Sec2}}

\subsection{Preliminaries\label{Sec2.1}}

In~a constant and homogeneous magnetic field background $\mathbf{B}$, the
potential $A_{0}\left( \mathbf{r}\right) $ generated by a stationary
point-like charge distribution $j^{\mu }\left( \mathbf{r}\right) =Ze\delta
_{0}^{\mu }\delta \left( \mathbf{r}\right) $, $\delta \left( \mathbf{r}%
\right) =\delta \left( x^{1}\right) \delta \left( x^{2}\right) \delta \left(
x^{3}\right) $ far from it takes the form of anisotropic Coulomb potential 
\cite{AdoGitSha16,ShaUso11,ChaSha12,GitSha12}%
\begin{equation}
A_{0}\left( \mathbf{r}\right) =\frac{Ze}{\sqrt{\varepsilon _{\perp }}\sqrt{%
\varepsilon _{\perp }x_{\parallel }^{2}+\varepsilon _{\parallel }\mathbf{x}%
_{\perp }^{2}}}\,,  \label{n1}
\end{equation}%
where $\varepsilon _{\perp }$\ and $\varepsilon _{\parallel }$\ are
dielectric permittivities, perpendicular to the external field and parallel
to it, respectively, and~$\varepsilon _{\perp }\neq \varepsilon _{\parallel }
$. These are functions of the background magnetic field\footnote{%
The~above expression must be understood in a special reference frame where
the background is purely magnetic and where the charge distribution is at
rest. Such a frame always exists, provided that the first field invariant $%
\mathfrak{F}=F_{\mu \nu }F^{\mu \nu }/4$ is positive and the second $%
\mathfrak{G}=F_{\mu \nu }\tilde{F}^{\mu \nu }/4$ identically zero.}. As~the
field of the charge in the remote region is slowly varying, these dielectric
permittivities are given by the entities associated with quantum
electrodynamics taken in the local field approximation (LCFA), i.e., the one
whose effective action is local in the sense that its Lagrangian $\mathfrak{L%
}$ is a function of the field invariants $\mathfrak{F},\mathfrak{G}$ only
(and not of space-time derivatives of them). Namely, the~permittivities are
expressed as $\varepsilon _{\perp }=1-\mathfrak{L}_{\mathfrak{F}}$\ and $%
\varepsilon _{\parallel }=\varepsilon _{\perp }+2\mathfrak{FL}_{\mathfrak{GG}%
}$\ in terms of the derivatives of the local Lagrangian $\mathfrak{L}$%
\begin{equation}
\mathfrak{L}_{\mathfrak{F}}=\left. \frac{\partial \mathfrak{L}}{\partial 
\mathfrak{F}}\right\vert _{\mathfrak{G}=0}\,,\ \ \mathfrak{L}_{\mathfrak{GG}%
}=\left. \frac{\partial ^{2}\mathfrak{L}}{\partial \mathfrak{G}^{2}}%
\right\vert _{\mathfrak{G}=0}\,.  \label{n2}
\end{equation}%

Because the background is purely magnetic, the~above coefficients must be
evaluated at $\mathfrak{G}=0$ after the partial differentiations. For our
purposes, the~one-loop approximation of the local Lagrangian $\mathfrak{L}$,
known as the Heisenberg--Euler Lagrangian~\cite{HeiEul36,Schwinger51}, is
fitting.\ However, it is important to note that Equation~(\ref{n1}) cannot be
blindly extrapolated onto the close vicinity of the charge, where a
Yukawa-like form of the potential (thus, quite distinct from (\ref{n1})) was
found beyond the local approximation in~\mbox{\cite{ShaUso07,ShaUso071,sadooghi}}.

The~leading large-field $b\gg 1$ asymptotic behavior for the dielectric
permittivities is%
\begin{equation}
\varepsilon _{\perp }\sim 1\,,\text{\ \ }\varepsilon _{\parallel }\sim 1+%
\frac{\alpha b}{3\pi }\,,  \label{As}
\end{equation}%
where $b=B/B_{\mathrm{cr}}$, $B_{\mathrm{cr}}=M^{2}c^{3}/e\hslash \approx
4.4\times 10^{13}\ \mathrm{G}$ is the critical Schwinger field, $e$ and $M$
are the electron charge and mass, and~$\alpha \approx \left( 137\right)
^{-1} $. It is this form that corresponds to the large-field behavior found
in~\cite{skobelev,skobelev2,melrose,Shabad88} for the polarization
tensor and exploited in~\cite{ShaUso07,ShaUso071}, while studying the
saturation phenomenon. We shall be basing this on the same Equation~(\ref{As}) in the
course of the present study of the saturation within LCFA.

The~following reservation is in order, however. The~next-to-leading terms of
the permittivities can be written with the help of the results in~\cite{heyl}
expressed in terms of Hurwitz Zeta functions as follows~\cite%
{AdoGitSha16,DitReu,DitGie,Elizalde,Kirsten,Dunne04,KarSha15}:%
\begin{eqnarray}
\varepsilon _{\perp } &=&1-\frac{\alpha }{2\pi }\left\{ \frac{2}{3}\ln 2b-%
\frac{1}{3}-\frac{1}{2b^{2}}\right.   \notag \\
&+&\left. \frac{1}{b}\left[ \ln \frac{\pi }{b}-2\ln \Gamma \left( \frac{1}{2b%
}\right) \right] +8\zeta ^{\prime }\left( -1,\frac{1}{2b}\right) \right\} \,,
\notag \\
\varepsilon _{\parallel } &=&1-\frac{\alpha }{3\pi }\left[ b+\ln 2b+\psi
\left( \frac{1}{2b}\right) \right] \,.  \label{ss2.1}
\end{eqnarray}%

Here, $\psi \left( a\right) =\Gamma ^{\prime }\left( a\right) /\Gamma \left(
a\right) $ denotes the Psi function~\cite{DLMF}, which in the large-field
limit acquires the asymptotic form%
\begin{equation*}
\psi \left( \frac{1}{2b}\right) =-2b-\gamma +O\left( b^{-1}\right) \,,\ \
b\rightarrow \infty \,,
\end{equation*}%
where $\gamma \approx 0.577$ denotes the Euler constant~\cite{DLMF}. The
logarithmic contribution $\varepsilon _{\perp }=1-\left( \alpha /3\pi
\right) \ln b$ into $\varepsilon _{\perp }$ may become essential under such
magnetic fields that the value $\left( \alpha /3\pi \right) \ln b$ becomes a
valuable portion of unity, say $1\%$, i.e., $b\gtrsim 10^{6}$. The~saturation
phenomenon we are going to consider here would be violated for those large
fields, but there is an important reason to refrain from including them into
consideration. The~point is that the logarithmic large-field terms in the
VP are known~\cite{Ritus3,Ritus31} to be associated with the
logarithmic large-momentum behavior that carries the
Fradkin--Landau--Pomeranchuk trouble~\cite{Fradkin,Fradkin2,Fradkin3} (also known as the lack of
asymptotic freedom) insurmountable within QED. In~the present context, the
logarithmic term $\left( \alpha /3\pi \right) \ln b$, if taken seriously,
would make the transverse dielectric permittivity negative at exponentially
large fields, which violates the causality, as~discussed in~\cite{ShaUso11}.
These conditions supply us with enough arguments to ignore next-to-leading
terms resulting from the large-field regime. Therefore, our working range of
magnetic fields is%
\begin{equation}
1\ll b< 10^{5}\,.  \label{range}
\end{equation}%

Note that these fields are smaller than/or of the order of the ones, whose
magnetic radius, expressed in terms of the electron Compton length $%
\lambdabar _{\mathrm{C}}=\hslash /Mc$ as $a_{\mathrm{H}}=\sqrt{\hslash c/eB}%
=\lambdabar _{\mathrm{C}}b^{-1/2}$, is of the order of/or smaller than the
electromagnetic radius of the\emph{\ }proton $\approx 2\lambdabar _{\mathrm{C%
}}\times 10^{-3}$~\cite{Codata}, so that the consideration of the hydrogen
atom might retain sense up to $b\lesssim 10^{5}$. Hence, no new restriction
on the range (\ref{range}) is produced.

Now we are in a position to explain preliminarily how the large screening
due to the linear growth of the longitudinal dielectric permittivity (\ref%
{As}) in the field range $b\gg 3\pi /\alpha =1.3\times 10^{3}$, which fits
the range (\ref{range}), is expected to provide finiteness of the
ground-state energy within LCFA. To this end, let us try to extrapolate (\ref%
{As}) taken in this regime as close to the charge as possible. The~smallest
transverse size of the electron Landau orbit in the Hydrogen atom in a
strong magnetic field is$\ a_{\mathrm{H}}$. Substituting this value for $%
\vert\mathbf{x}_{\perp}\vert$ into (\ref{As}), we~obtain the potential%
\begin{equation}
\left. A_{0}\left( \mathbf{r}\right) \right\vert _{\vert\mathbf{x}_{\perp}\vert=a_{\mathrm{H}}}\equiv A_{0}\left( x_{\parallel }\right) =\frac{Ze}{\sqrt{x_{\parallel }^{2}+\varepsilon _{\parallel }a_{\mathrm{H}}^{2}}}\,.
\label{Pot}
\end{equation}%

If it were not for the radiative corrections, we~would have $\varepsilon
_{\parallel }=1$ and this potential in the large field limit would be the
singular Coulomb one $Ze/x_{\parallel }$, since $a_{\mathrm{H}%
}^{2}\rightarrow 0$. With such a singularity, the~one-dimensional Schr\"{o}%
dinger equation governing the degree of freedom parallel to the magnetic
field yields an unlimitedly deep energy level according to~\cite{EllLou60,EllLou6066}.
On the contrary, with the linearly growing longitudinal dielectric
permittivity $\varepsilon _{\parallel }=\left( \alpha /3\pi \right) b$, the
potential, as~a function of the longitudinal coordinate, remains ever-finite%
\begin{equation*}
A_{0}\left( x_{\parallel }\right) =\frac{Ze}{\sqrt{x_{\parallel }^{2}+\left(
\alpha /3\pi \right) \lambdabar _{\mathrm{C}}^{2}}}\,.
\end{equation*}%

This~excludes unrestrictedness of the energy level. Moreover, this~fact may
justify the performed extrapolation of the potential into the short-range
domain, where only the polarization tensor beyond LCFA is expected to be
valid. To forestall our result for the ground state energy, we~note that,
numerically, it closely coincides with the value calculated in \cite%
{MacVys11} beyond LCFA, where the large screening provides the saturation
via a different mechanism. Namely, it replaces the Coulomb singularity $%
Ze/x_{\parallel }$ by the Dirac delta-function $2.18\times \left( Ze/2\pi
\right) \delta \left( x_{\parallel }\right) $~\cite{ShaUso07,ShaUso071}, to which the
Yukawa potential turns in the limit $b\gg 3\pi /\alpha $.

\subsection{Non-Relativistic Wave Equations, Diagonal Approximation and
Effective Potential\label{Sec2.2}}

The~nonrelativistic electron in the modified Coulomb field $V\left( \mathbf{r%
}\right) =-eA_{0}\left( \mathbf{r}\right) $\ plus a constant magnetic field $%
\mathbf{B}$ will be described here by the Pauli Hamiltonian $\hat{\mathbb{H}}
$,%
\begin{equation}
\hat{\mathbb{H}}=\frac{1}{2M}\left( \boldsymbol{\sigma }\mathbf{\hat{P}}%
\right) ^{2}-eA_{0}\left( \mathbf{r}\right) =\hat{\mathbb{H}}^{\left(
0\right) }+V\left( \mathbf{r}\right) \,,  \label{nr0}
\end{equation}%
where $\mathbf{\hat{P}}=\mathbf{\hat{p}}+\left( e/c\right) \mathbf{A}%
=-i\hslash \boldsymbol{\nabla }+\left( e/2c\right) \left[ \mathbf{B}\times 
\mathbf{r}\right] $, $\boldsymbol{\sigma }$ are Pauli matrices, $M$ is the
electron mass and $e$ its charge, in~absolute value. The~Hamiltonian $\hat{%
\mathbb{H}}^{\left( 0\right) }$ describes the electron motion and its
interaction with the external magnetic field $\mathbf{B}$%
\begin{equation}
\hat{\mathbb{H}}^{\left( 0\right) }=\frac{1}{2M}\left\{ -\hslash ^{2}%
\boldsymbol{\nabla }^{2}+\frac{e}{c}\left( \mathbf{B\hat{L}}\right) +\frac{%
e^{2}}{4c^{2}}\left[ \mathbf{B}^{2}r^{2}-\left( \mathbf{Br}\right) ^{2}%
\right] +\frac{2e}{c}\left( \mathbf{B\hat{s}}\right) \right\} \,,  \label{n3}
\end{equation}%
where $\mathbf{\hat{L}}=\mathbf{r}\times \mathbf{\hat{p},}$ and $\mathbf{%
\hat{s}}=\hslash \boldsymbol{\sigma }/2$ denotes the ordinary angular
momentum and spin operators, respectively. Because of its cylindric
symmetry, $\hat{\mathbb{H}}^{\left( 0\right) }$ splits into two parts $\hat{%
\mathbb{H}}^{\left( 0\right) }=\hat{\mathbb{H}}_{\perp }^{\left( 0\right) }+%
\hat{\mathbb{H}}_{\parallel }^{\left( 0\right) }$,\ longitudinal, $\hat{%
\mathbb{H}}_{\parallel }^{\left( 0\right) }=-\left( \hslash ^{2}/2M\right)
\partial ^{2}/\partial x_{\parallel }^{2},$ and transverse, $\hat{\mathbb{H}}%
_{\perp }^{\left( 0\right) }$, with respect to $\mathbf{B}$ (positively
oriented along the $z$-axis from now on, $\mathbf{B}=B\mathbf{\hat{z}}$). $%
\hat{\mathbb{H}}_{\perp }^{\left( 0\right) }$ is the Landau Hamiltonian
describing the motion on the $xy$ plane, whose eigenvalues $\mathcal{E}%
_{n_{\rho }m\sigma }^{\left( 0\right) }$ and eigenvectors $\Phi _{n_{\rho
}m\sigma }\left( \rho ,\varphi \right) $ are $\mathcal{E}_{n_{\rho }m\sigma
}^{\left( 0\right) }=\left( \hslash eB/Mc\right) \left[ n_{\rho }+\left(
\left\vert m\right\vert +m+1+\sigma \right) /2\right] $ and $\Phi _{n_{\rho
}m\sigma }\left( \rho ,\varphi \right) =\left( 2\pi \right)
^{-1/2}e^{im\varphi }R_{n_{\rho }m}\left( \rho \right) v_{\sigma }$,
respectively. Here, $\rho =\sqrt{x^{2}+y^{2}}$ and $\varphi =\arctan \left(
y/x\right) $ are polar variables while $v_{\sigma }$ eigen-spinors of $%
\sigma _{z}$, $\sigma _{z}v_{\sigma }=\sigma v_{\sigma }$, $v_{\sigma
}^{\dagger }v_{\sigma ^{\prime }}=\delta _{\sigma \sigma ^{\prime }}$, $%
\sigma =\pm 1$. The~quantum number\textrm{\ }$m=0,\pm 1,\pm 2,...$
corresponds to eigenvalues of the projection of the angular momentum
operator on the $z$-axis while $n_{\rho }=0,1,2,...$ to the radial quantum
number resulting from the transverse motion, respectively. The~explicit form
of the radial functions $R_{n_{\rho }m}\left( \rho \right) $ reads%
\begin{equation*}
R_{n_{\rho }m}\left( \rho \right) =\sqrt{\frac{\left( \left\vert
m\right\vert +n_{\rho }\right) !}{2^{\left\vert m\right\vert }n_{\rho }!}}%
\frac{e^{-\rho ^{2}/4a_{\mathrm{H}}^{2}}\rho ^{\left\vert m\right\vert }}{a_{%
\mathrm{H}}^{1+\left\vert m\right\vert }\left\vert m\right\vert !}\Phi
\left( -n_{\rho },\left\vert m\right\vert +1;\frac{\rho ^{2}}{2a_{\mathrm{H}%
}^{2}}\right) \,,
\end{equation*}%
in which $\Phi \left( a,c;y\right) $ denotes the confluent hypergeometric
function (CHF)~\cite{Erdelyi}.

We are primarily interested in the lower state energies of the stationary
Schr\"{o}dinger equation%
\begin{equation}
\hat{\mathbb{H}}\Psi \left( \mathbf{r}\right) =\mathcal{E}\Psi \left( 
\mathbf{r}\right) \,,  \label{hy10}
\end{equation}%
specialized to cases where $V\left( \mathbf{r}\right) $ can be treated as a
perturbation over the magnetic background. It is noteworthy that even in the
approximation where VP is ignored, the~problem cannot be
straightforwardly solved because the spherical symmetry of the ordinary
Coulomb field $Ze/r$ and the cylindrical symmetry from the magnetic field
compete with one another, preventing a complete separation of variables of
the Hamiltonian (\ref{nr0}) for magnetic fields of arbitrary strength; see,
e.g., Refs.~\cite{RuderBook,FriWin89}, and~pertinent references therein. For
sufficiently strong magnetic backgrounds, the~usual approach to the problem
consists of expanding the wave function in Landau states%
\begin{equation}
\Psi \left( \mathbf{r}\right) =\sum_{n_{\rho }^{\prime }m^{\prime }\sigma
^{\prime }}\chi _{n_{\rho }^{\prime }m^{\prime }}\left( z\right) \Phi
_{n_{\rho }^{\prime }m^{\prime }\sigma ^{\prime }}\left( \rho ,\varphi
\right) \,,  \label{hy11}
\end{equation}%
where $\chi _{n_{\rho }m}\left( z\right) $ are expansion functions to be
determined. For convenience of notations, we~henceforward use $z$ for the
longitudinal coordinate, $z=x_{\parallel }$. Plugging the expansion (\ref%
{hy11}) into Equation~(\ref{hy10}) and using the orthogonality of the Landau
states, the~expansion functions are found to be solutions of the system of
differential equations%
\begin{equation}
\left( -\frac{\hslash ^{2}}{2M}\frac{d^{2}}{dz^{2}}+\lambda _{n_{\rho
}m\sigma }^{2}\right) \chi _{n_{\rho }m}\left( z\right) +\sum_{n_{\rho
}^{\prime }m^{\prime }\sigma ^{\prime }}\mathbb{U}_{n_{\rho }n_{\rho
}^{\prime }mm^{\prime }\sigma \sigma ^{\prime }}\left( z\right) \chi
_{n_{\rho }^{\prime }m^{\prime }}\left( z\right) =0\,,  \label{ad1.4}
\end{equation}%
provided the binding energies $\lambda _{n_{\rho }m\sigma }^{2}$, $\mathcal{E%
}$ and $\mathcal{E}_{n_{\rho }m\sigma }^{\left( 0\right) }$ are related as%
\begin{equation}
\lambda _{n_{\rho }m\sigma }^{2}=\mathcal{E}_{n_{\rho }m\sigma }^{\left(
0\right) }-\mathcal{E}\,.  \label{ad1.4b}
\end{equation}%
and $\mathbb{U}_{n_{\rho }n_{\rho }^{\prime }mm^{\prime }\sigma \sigma
^{\prime }}\left( z\right) $ ---the \textit{effective potential} \cite%
{SnySch39,Landauvol3,RuderBook} ---is defined as%
\begin{equation}
\mathbb{U}_{n_{\rho }n_{\rho }^{\prime }mm^{\prime }\sigma \sigma ^{\prime
}}\left( z\right) =\int_{0}^{2\pi }d\varphi \int_{0}^{\infty }d\rho \rho %
\left[ \Phi _{n_{\rho }m\sigma }^{\ast }\left( \rho ,\varphi \right) V\left( 
\mathbf{r}\right) \Phi _{n_{\rho }^{\prime }m^{\prime }\sigma ^{\prime
}}\left( \rho ,\varphi \right) \right] \,.  \label{ad1.4c}
\end{equation}%

It should be noted that the effective potential is diagonal in the spinning
and angular-momentum quantum numbers%
\begin{equation}
\mathbb{U}_{n_{\rho }n_{\rho }^{\prime }mm^{\prime }\sigma \sigma ^{\prime
}}\left( z\right) =\delta _{mm^{\prime }}\delta _{\sigma \sigma ^{\prime }}%
\mathbb{U}_{n_{\rho }n_{\rho }^{\prime }}^{\left\vert m\right\vert }\left(
z\right) \,,\ \ \mathbb{U}_{n_{\rho }n_{\rho }^{\prime }}^{\left\vert
m\right\vert }\left( z\right) =-\frac{Ze^{2}}{\sqrt{\varepsilon _{\perp }}}%
\int_{0}^{\infty }d\rho \rho \frac{R_{n_{\rho }m}^{\ast }\left( \rho \right)
R_{n_{\rho }^{\prime }m}\left( \rho \right) }{\sqrt{\varepsilon _{\perp
}z^{2}+\varepsilon _{\parallel }\rho ^{2}}}\,,  \label{ad1.4d}
\end{equation}%
on account of the commutativity between $V\left( \mathbf{r}\right) $, $\hat{L%
}_{z}$, and~$\hat{s}_{z}=\hslash \sigma _{z}/2$.

In~general, Equation~(\ref{ad1.4}) is a coupled system of differential equations
for the functions $\chi _{n_{\rho }m}\left( z\right) $ and the binding
energies $\lambda _{n_{\rho }m\sigma }^{2}$ which can be treated numerically
using Hartree--Fock methods, for instance~\cite{RuderBook}. However,
restricting the consideration to strong magnetic fields supplies us with the
possibility of treating Equation~(\ref{ad1.4}) analytically within the adiabatic
approximation~\cite{SnySch39}, in~which one neglects contributions from
non-diagonal elements of the effective potential $\mathbb{U}_{n_{\rho
}n_{\rho }^{\prime }}^{\left\vert m\right\vert }\left( z\right) =\delta
_{n_{\rho }n_{\rho }^{\prime }}\mathbb{U}_{n_{\rho }}^{\left\vert
m\right\vert }\left( z\right) $, reducing the wave function (\ref{hy11}) to $%
\Psi \left( \mathbf{r}\right) =\chi _{n_{\rho }m}\left( z\right) \Phi
_{n_{\rho }m\sigma }\left( \rho ,\varphi \right) $ and the system (\ref%
{ad1.4}) to the one-dimensional motion%
\begin{equation}
\left[ -\frac{\hslash ^{2}}{2M}\frac{d^{2}}{dz^{2}}+\lambda _{n_{\rho
}m\sigma }^{2}+\mathbb{U}_{n_{\rho }}^{\left\vert m\right\vert }\left(
z\right) \right] \chi _{n_{\rho }m\sigma }\left( z\right) =0\,,
\label{ad1.13}
\end{equation}%
see, e.g., Refs. \cite%
{PopKar03,PopKar12,PopKar14,RuderBook,SimVit78,WunRud80,WunRud80b,WunRudHer81,Fri82,Rosner82}%
. This~approximation has been scrutinized over the years, in~particular, it
was numerically established that it becomes sufficiently accurate for
magnetic fields satisfying $\mathcal{B}\equiv B/B_{\mathrm{a}}>10^{3}$,\
where $B_{\mathrm{a}}=\alpha ^{2}B_{\mathrm{cr}}=M^{2}e^{3}c/\hslash
^{3}\approx 2.4\times 10^{9}\ \mathrm{G}$ is a reference magnetic field
strength for atomic systems\footnote{%
The~reference magnetic field $B_{\mathrm{a}}$ is chosen such that the
corresponding oscillator energy $\hslash \omega _{\mathrm{a}}/2=\hslash eB_{%
\mathrm{a}}/2Mc$ equals the Rydberg energy $\mathrm{Ry}=Me^{4}/2\hslash ^{2}$%
.}~\cite{WunRud80,RuderBook,WunRud80b}. Rephrasing this condition in terms
of the ratio $b=\alpha ^{2}\mathcal{B}$, it corresponds to $b>10^{-1}$. This
condition is met within our working range (\ref{range}).

To estimate the energy of lower levels (more precisely, the~lowest Landau
level band (LLL), specified by $n_{\rho }=0$, $\,m=-\left\vert m\right\vert
=0,1,2,...,$ $\sigma =-1$), we~first note that the wave functions $\chi
_{n_{\rho }m\sigma }\left( z\right) $ are either even or odd with respect to 
$z$ owing to the symmetry of (\ref{n1}) under space reflection, $A_{0}\left( 
\mathbf{r}\right) =A_{0}\left( -\mathbf{r}\right) $. Hence, we~follow the
procedure proposed by Karnakov and Popov (KP) to obtain a spectrum equation
for \textit{even} levels~\cite{PopKar03,PopKar12,PopKar14}, which consists
of solving the problem in three steps: first at short distances, where Equation~(%
\ref{ad1.13}) can be solved perturbatively, second at sufficiently large
distances, where it is exactly solvable, and~third equating the logarithmic
derivative of the solutions at the overlapping region. To this aim, we~first
calculate the corresponding effective potentials $\mathbb{U}_{0}^{\left\vert
m\right\vert }\left( z\right) $ which, according to Equation~(\ref{ad1.4d}), are
confluent hypergeometric functions $\Psi \left( a,c;y\right) $~\cite{Erdelyi}%
,%
\begin{equation}
\mathbb{U}_{0}^{\left\vert m\right\vert }\left( z\right) =-\frac{Ze^{2}}{%
\sqrt{2a_{\mathrm{H}}^{2}\varepsilon _{\perp }\varepsilon _{\parallel }}}%
\Psi \left( \frac{1}{2},\frac{1}{2}-\left\vert m\right\vert ;\mathcal{X}%
^{2}\right) \,,\ \ \mathcal{X}^{2}=\frac{\varepsilon _{\perp }}{\varepsilon
_{\parallel }}\frac{z^{2}}{2a_{\mathrm{H}}^{2}}\,.  \label{ad1.15}
\end{equation}%

\begin{figure}[h!]
\centering
\includegraphics[scale=0.55]{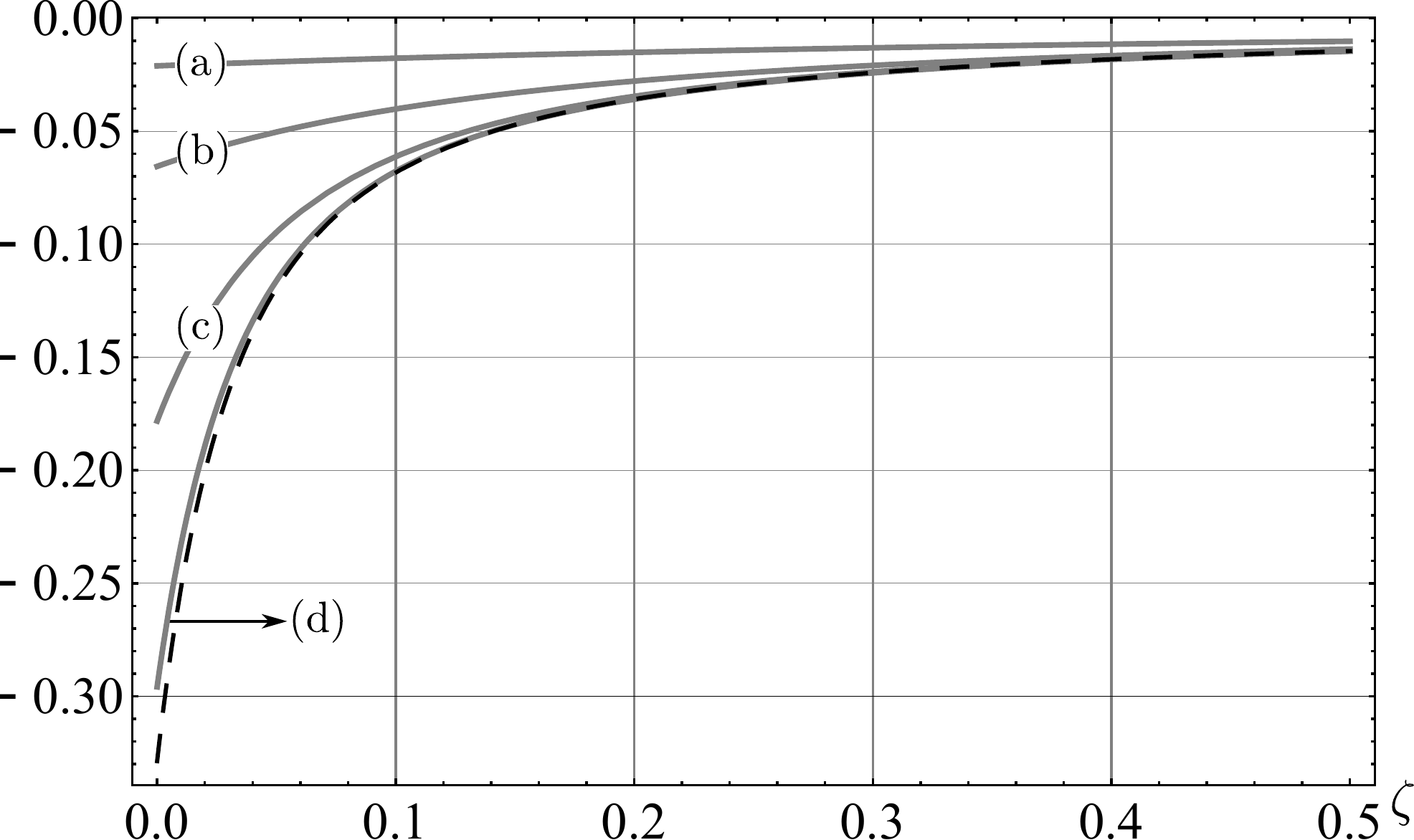}
\caption{Condensation of the effective potentials (\protect
\ref{ad1.15}) onto the saturation curve (\protect\ref{ss3}) ($Z=1$). All~potentials, given in units of the electron rest energy $Mc^{2}$, are
functions of the distance $\protect\zeta =z/\lambdabar _{\mathrm{C}}$ and
are labeled by the values of magnetic fields of the following amplitudes:
(\textbf{a}) $\mathcal{B}=10^{5}$ ($b\approx 5.3$), (\textbf{b})~$\mathcal{B}=10^{6}$ ($%
b\approx 5.3\times 10$), (c) $\mathcal{B}=10^{7}$ ($b\approx 5.3\times 10^{2}
$), and~(\textbf{d}) $\mathcal{B}=10^{8}$ ($b\approx 5.3\times 10^{3}$). Their
limiting values at $z=0$, are, approximately, $-0.0211$ for (\textbf{a}), $-0.0656$
for (\textbf{b}), $-0.1782$ for (\textbf{c}), and~$-0.2958$\ for~(\textbf{d}).}
\label{F0}
\end{figure}

\begin{figure}[h!]
\centering
\includegraphics[scale=0.38]{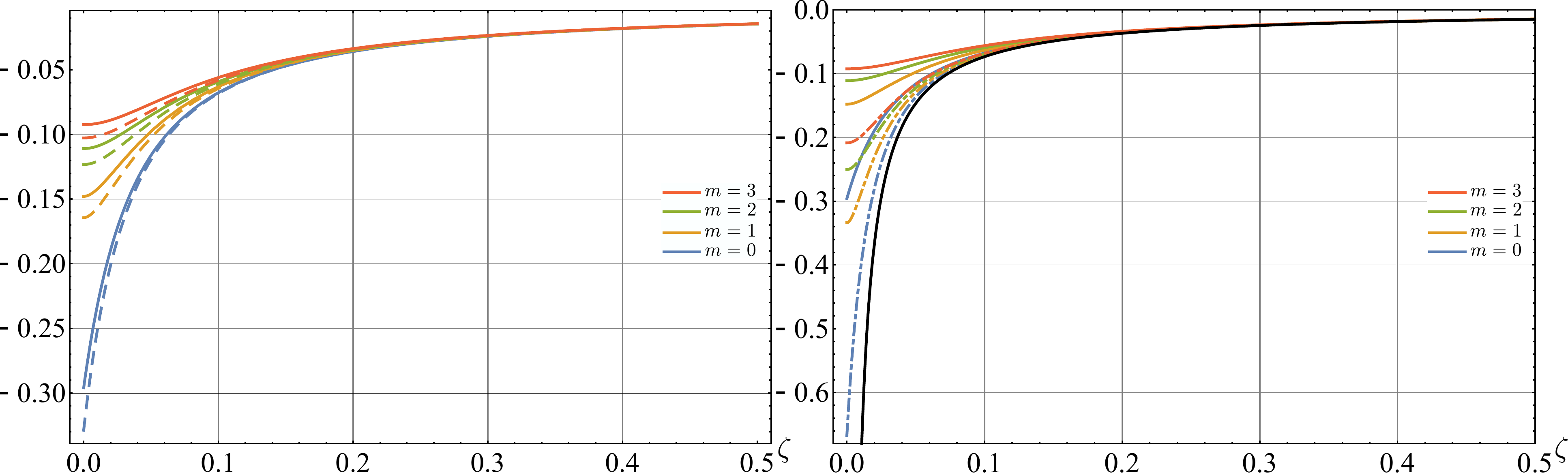}
\caption{(color online) Effective potentials (in units of the electron rest
energy $Mc^{2}$) as functions of the distance $\protect\zeta =z/\lambdabar _{\mathrm{c}}$ for $\mathcal{B}=10^{8}$. The~left panel displays the effective
potentials (\protect\ref{ad1.15}) (solid lines) and their corresponding
saturation curves (\protect\ref{ss3}) (dashed lines). In~the right panel,
the solid colored lines are the same as in the left panel while the
dot-dashed lines refer to effective potentials neglecting vacuum
polarization, given by Equation~(\protect\ref{ad1.15}) with $\protect\varepsilon %
_{\perp }=\protect\varepsilon _{\parallel }=1$. The~solid black line
corresponds to the large-distance Coulomb potential (\protect\ref{ad1.15c}).}
\label{F1}
\end{figure}

For the purposes of the present work, the~permittivities $\varepsilon
_{\perp }$\ and $\varepsilon _{\parallel }$\ will be understood mostly as
their leading asymptotic limit (\ref{As}). We find it useful, however, to
keep them explicitly for completeness in equations below. From asymptotic
properties of CHF with large argument and limiting values as $z\rightarrow 0$
\cite{DLMF}, the~effective potentials (\ref{ad1.15}) are Coulombian at large
distances%
\begin{equation}
\mathbb{U}_{0}^{\left\vert m\right\vert }\left( z\right) \sim \mathbb{U}_{%
\mathrm{C}}\left( z\right) =-\frac{Ze^{2}}{\varepsilon _{\perp }}\frac{1}{%
\left\vert z\right\vert }\,,\ \ z^{2}\rightarrow \infty \,,  \label{ad1.15c}
\end{equation}%
and regular at the origin,%
\begin{equation}
\mathbb{U}_{0}^{\left\vert m\right\vert }\left( 0\right) =-\frac{Ze^{2}}{%
\sqrt{2a_{\mathrm{H}}^{2}\varepsilon _{\perp }\varepsilon _{\parallel }}}%
\frac{\Gamma \left( \left\vert m\right\vert +1/2\right) }{\Gamma \left(
\left\vert m\right\vert +1\right) }\,,  \label{ad1.15b}
\end{equation}%
whose minimum corresponds to the lowest value for $m$, $\mathbb{U}%
_{0}^{0}\left( 0\right) =-Ze^{2}\sqrt{\pi }/\sqrt{2a_{\mathrm{H}%
}^{2}\varepsilon _{\perp }\varepsilon _{\parallel }}$. This~value is finite
and, up to a numerical factor, is in agreement with the potential (\ref{Pot}%
) taken in $x_{\parallel }=0$. Moreover, the~effective potential curves (\ref%
{ad1.15}) %Please check that intended meaning has been retained. [REPLY: OK, we agree with this correction]
condense towards the prescribed curve,%
\begin{equation}
\mathbb{U}_{0}^{\left\vert m\right\vert }\left( z\right) \sim \mathbb{U}_{%
\mathrm{sat}}^{\left\vert m\right\vert }\left( z\right) =ZMc^{2}\sqrt{\frac{%
3\pi \alpha }{2}}\Psi \left( \frac{1}{2},\frac{1}{2}-\left\vert m\right\vert
;\frac{3\pi }{2\alpha }\zeta ^{2}\right) \,,\,\,\zeta =\frac{z}{\lambdabar _{%
\text{\textrm{C}}}}\,,  \label{ss3}
\end{equation}%
as the magnetic field grows. To see this, note that in the range of magnetic
fields under consideration, the~asymptotic behavior of the dielectric
permittivities (\ref{ss2.1}) is given by Equation~(\ref{As}). In~cases where
VP is ignored~\cite{PopKar03,PopKar12,PopKar14}, the
effective potentials do not condense, and~their minima decrease indefinitely
as the magnetic field grows, as~can be seen from Equation~(\ref{ad1.15b}) by
setting $\varepsilon _{\perp }=\varepsilon _{\parallel }=1$. For
illustrative purposes, we~present in Figure~\ref{F0} the effective potentials (%
\ref{ad1.15}) (with $m=0$) for some values of the external field and its
corresponding saturation curve, given by Equation~(\ref{ss3}). Moreover, we
supply the first picture with curves corresponding to different values of $m$
but a fixed value of the external field, $\mathcal{B}=10^{8}$ ($b\approx
5.3\times 10^{3}$). This~is illustrated by the left panel of Figure~\ref{F1}.
In~the right panel, we~compare the potentials (\ref{ad1.15}) (color solid
lines) with the large-distance Coulomb potential (\ref{ad1.15c}) (black
line) and with potentials whose VP effects are neglected
(color dashed lines), Equation~(\ref{ad1.15}) with $\varepsilon _{\perp
}=\varepsilon _{\parallel }=1$\ in~it.

\unskip

\subsection{Finding the spectrum\label{Sec2.3}}

Now we are in a position to consider the Schr\"{o}dinger equation at short
distances $z\in \left[ 0,\ell \right] $ (with $\ell $ sufficiently small).
Following~\cite{Landauvol3,PopKar03,PopKar12,PopKar14}, we~seek for a
perturbative solution in the so-called shallow-well approximation,
which can be applied whenever the effective potentials are sufficiently
\textquotedblleft shallow\textquotedblright\ according to the condition%
\begin{equation}
M\left\vert \mathbb{U}_{0}^{\left\vert m\right\vert }\left( z\right)
\right\vert \frac{\ell ^{2}}{\hslash ^{2}}\ll 1\,,\ \ z\in \left[ 0,\ell %
\right] \,.  \label{ad1.14}
\end{equation}%

This~approximation was first considered in one-dimensional problems of
quantum mechanics~\cite{Landauvol3}, in~which $\ell $ stands for the support
of the potential well. Because the extent of the effective potential~(\ref%
{ad1.15}) is infinite, it is necessary to choose an appropriately small
value to $\ell $ such that the potential be sufficiently shallow according
to (\ref{ad1.14}), but also such that for $z>\ell $ the potential (\ref%
{ad1.15}) might be considered Coulombian (\ref{ad1.15c}) with sufficiently
high accuracy. We assume here that the condition (\ref{ad1.14}) is satisfied
and postpone the discussion about its validity to the next Section \ref{Sec3}%
, where numerical calculations are discussed in detail.

At this point, it is important to mention that for every pair of quantum
numbers $n_{\rho },m$, there is an infinite number of longitudinal quantum
numbers $\nu $ labeling the longitudinal states\footnote{%
Hereafter, we~drop the label $\sigma $ from the wave functions $\chi
_{n_{\rho }m\sigma }\left( z\right) $ for simplicity.} $\chi _{n_{\rho
}m}^{\nu }\left( z\right) $ such that a complete description of states and
energy shifts $\lambda _{n_{\rho }m\sigma }^{2}$ is given by the set of
quantum numbers $\nu ,n_{\rho },m,\sigma $. Thus, performing a change of
variables $\xi =z/a_{\mathrm{B}}$, $a_{\mathrm{B}}=\hslash ^{2}/Me^{2}$, Equation
(\ref{ad1.13}) for the LLL~reads%
\begin{equation}
\frac{d^{2}}{d\xi ^{2}}\chi _{0m}^{\nu }\left( \xi \right) =\left( \omega
_{\nu }^{2}+\frac{\mathbb{U}_{0}^{\left\vert m\right\vert }\left( \xi
\right) }{\mathrm{Ry}}\right) \chi _{0m}^{\nu }\left( \xi \right) \,,
\label{ad1.17}
\end{equation}%
where $\mathrm{Ry}=Me^{4}/2\hslash ^{2}$ is the Rydberg energy and $\omega
_{\nu }^{2}=\lambda _{\nu 0m-1}^{2}/\mathrm{Ry}$. In~view of the condition (%
\ref{ad1.14}) and the restriction to even levels, the~wave function of
longitudinal states $\chi _{0m}^{\nu }\left( z\right) $ varies slowly within
the interval $\left[ 0,\ell \right] $ ($\chi _{0m}\left( \ell \right)
\approx \chi _{0m}\left( 0\right) =\mathrm{const}$.) and the ground-state
binding energies are negligible in comparison to the potential, $\lambda
_{\nu 00-1}\ll \left\vert \mathbb{U}_{0}^{\left\vert m\right\vert }\left(
z\right) \right\vert $ for $z\in \left[ 0,\ell \right] $. As~a result, one
can integrate Equation~(\ref{ad1.17}) to show that the logarithmic derivative has
the form%
\begin{equation}
\begin{array}{lll}
\dfrac{\chi _{0m}^{\nu\prime}\left( \xi \right) }{\chi _{0m}^{\nu }\left(
\xi \right) } &\approx &-\dfrac{4Z}{\varepsilon _{\perp }\left( 2a_{\mathrm{H}%
}^{2}\right) ^{\left\vert m\right\vert +1}\left\vert m\right\vert !}%
\int_{0}^{\infty }d\rho \rho \rho ^{2\left\vert m\right\vert }e^{-\rho
^{2}/2a_{\mathrm{H}}^{2}}\int_{0}^{\xi }\dfrac{d\tilde{\xi}}{\sqrt{\tilde{\xi}%
^{2}+\varepsilon _{\parallel }\rho ^{2}/\varepsilon _{\perp }a_{\mathrm{B}%
}^{2}}}   \\[2ex]
&\approx &-\dfrac{Z}{\varepsilon _{\perp }}\left[ \ln 4\mathcal{X}^{2}-\psi
\left( \left\vert m\right\vert +1\right) \right] \,.  \label{ad1.18}
\end{array}\end{equation}

To obtain Equation~(\ref{ad1.18}), we~used the boundary conditions $\chi
_{0m}^{\nu }\left( 0\right) =1$, $\chi _{0m}^{\nu \prime }\left( 0\right) =0$
and the approximation%
\begin{equation}
\int_{0}^{\xi }dx\left( x^{2}+\frac{\varepsilon _{\parallel }\rho ^{2}}{%
\varepsilon _{\perp }a_{\mathrm{B}}^{2}}\right) ^{-1/2}\approx \ln \frac{%
2\xi }{\sqrt{\varepsilon _{\parallel }\rho ^{2}/\varepsilon _{\perp }a_{%
\mathrm{B}}^{2}}}\,,  \label{ad1.20}
\end{equation}%
provided $\xi ^{2}\gg \varepsilon _{\parallel }\rho ^{2}/\varepsilon _{\perp
}a_{\mathrm{B}}^{2}$. This~condition, which will be clarified in the next
section, is met for magnetic fields within the range (\ref{range}) and for $z
$ larger than the characteristic length of the transverse motion\footnote{%
The~dominant contribution to the integral over $\rho $ in (\ref{ad1.18})
comes from values beneath the Landau radius $\rho \lesssim $ $a_{\mathrm{H}}$%
, due to the exponential damping from the wave functions $R_{0m}\left( \rho
\right) $.}. As~for large distances, one can substitute the
approximation (\ref{ad1.15c}) into Equation~(\ref{ad1.17}) to obtain a Schr\"{o}%
dinger equation with one-dimensional Coulomb-like potential \cite%
{EllLou60,EllLou6066,PopKar03,PopKar12,PopKar14}. The~corresponding solution regular at
infinity is the Whittaker function $\chi _{0m}^{\nu }\left( \xi \right)
=W_{\kappa ,1/2}\left( 2\omega _{\nu }\xi \right) $, in~which $\kappa
=Z/\omega _{\nu }\varepsilon _{\perp }$. For sufficiently small arguments,
such a solution behaves according to Equation~(3.5') in Ref.~\cite{PopKar14} (see~also Equation~(13.14.17) in~\cite{DLMF}). Computing the corresponding derivative,
we obtain the following approximate expression for the logarithmic
derivative,%
\begin{equation}
\frac{\chi _{0m}^{\prime \nu }\left( \xi \right) }{\chi _{0m}^{\nu }\left(
\xi \right) }\approx -\omega _{\nu }-2\omega _{\nu }\kappa \left[ \ln
2\omega _{\nu }\xi +\psi \left( 1-\kappa \right) +2\gamma \right] \,.
\label{ad1.19}
\end{equation}%

Matching Equations (\ref{ad1.18}) and (\ref{ad1.19}) we obtain a modified KP equation for even levels\footnote{Apart from the usual gauge invariance of the eigenvalues of the Schr\"{o}dinger equation with external potentials, it is worth noticing that this symmetry is preserved in Equation~(\ref{ad2}), since the dielectric permittivities depend only on field intensities, and, besides, they are obtained from the gauge-invariant (4-transverse) polarization tensor, like in~\cite{ShaUso07,ShaUso071,BatSha71,BatSha712,BatSha713,BatSha714,BatSha715,Shabad88,ShaUso11,FerSan19}.}%
\begin{equation}
\ln \mathcal{B}=\frac{\varepsilon _{\perp }\omega _{\nu }}{Z}+2\ln \omega
_{\nu }+2\psi \left( 1-\frac{Z}{\varepsilon _{\perp }\omega _{\nu }}\right)
+4\gamma +\ln 2+\psi \left( \left\vert m\right\vert +1\right) +\ln \left( 
\frac{\varepsilon _{\parallel }}{\varepsilon _{\perp }}\right) \,.
\label{ad2}
\end{equation}%

The~above equation coincides with the KP equation in the regime without
radiative corrections.

\section{Results\label{Sec3}}

To justify the spectrum equation for the lowest energy levels, we~must
ascertain whether the conditions under which the approximations considered
before are satisfied. More precisely, the~modified KP equation (\ref{ad2})
relies on the assumption that the shallow-well condition (\ref{ad1.14}) is
satisfied at small distances and the fact that the effective potential is
Coulombian at large distances (\ref{ad1.15c}), with~sufficiently high
accuracy.

Now, we~proceed to study the shallow-well condition (\ref{ad1.14}), as~a
function of the magnetic field $B$ and the distance from the origin,
assuming fixed values for $\ell $. Letting $\ell =K\lambdabar _{\mathrm{C}}$%
, with $K$ an arbitrary positive number, we~conveniently rewrite the
left-hand side of the condition (\ref{ad1.14}) as%
\begin{equation}
\Xi ^{\left\vert m\right\vert }\left( \mathcal{B},\zeta \right) =Z\alpha ^{2}%
\sqrt{\frac{\mathcal{B}}{2\left\vert \varepsilon _{\perp }\varepsilon
_{\parallel }\right\vert }}\left\vert \Psi \left( \frac{1}{2},\frac{1}{2}%
-\left\vert m\right\vert ;\frac{\varepsilon _{\perp }}{\varepsilon
_{\parallel }}\frac{\mathcal{B}}{2}\alpha ^{2}\zeta ^{2}\right) \right\vert
K^{2}\,.  \label{n3.1}
\end{equation}

Choosing $K=1.5$ and using Equation (\ref{ss2.1}), one can calculate the
coefficient $\Xi ^{\left\vert m\right\vert }\left( \mathcal{B},\zeta \right) 
$ numerically to conclude that $0.01\lesssim \Xi ^{0}\left( 10^{5},\zeta
\right) \lesssim 0.047$ for weaker fields while $0.011\lesssim \Xi
^{0}\left( 10^{9},\zeta \right) \lesssim 0.733$ for stronger fields,
admitting distances within the interval $z/\lambdabar _{\mathrm{C}}=\zeta
\in \left[ 0,1.5\right] $. As~for larger supports, say~$\ell =3\lambdabar _{%
\mathrm{C}}$, the~condition (\ref{ad1.14}) is preserved for weaker fields $%
0.2\lesssim \Xi ^{0}\left( 10^{5},\zeta \right) \lesssim 0.022$ but violated
for strong fields, since $0.022\lesssim \Xi ^{0}\left( 10^{9},\zeta \right)
\lesssim 2.93$, assuming $\zeta \in \left[ 0,3\right] $. The~condition (\ref%
{ad1.14}) is improved as $m$ grows since the larger the $m$, the~shallower
the effective potential. For example, $0.01\lesssim \Xi ^{1}\left(
10^{5},\zeta \right) \lesssim 0.024$ and $0.01\lesssim \Xi ^{1}\left(
10^{9},\zeta \right) \lesssim 0.367$, assuming $\zeta \in \left[ 0,1.5\right]
$. Increasing the support to $\ell =3\lambdabar _{\mathrm{C}}$, we~find $%
0.021\lesssim \Xi ^{1}\left( 10^{5},\zeta \right) \lesssim 0.1$ and $%
0.02\lesssim \Xi ^{1}\left( 10^{9},\zeta \right) \lesssim 1.47$\ for $\zeta
\in \left[ 0,3\right] $. Based on these results, we~conclude that there is a
range of values to $\ell $ for which the condition (\ref{ad1.14}) is
fulfilled for all values of the magnetic field under consideration. Next,
one has to verify to what extent the effective potential (\ref{ad1.15}) is
Coulombian (\ref{ad1.15c}) for $z$ slightly larger than $\ell $. This~can be
done, for example, analyzing the ratio $R^{\left( \left\vert m\right\vert
\right) }\left( z\right) =\mathbb{U}_{0}^{\left\vert m\right\vert }\left(
z\right) /\mathbb{U}_{\mathrm{C}}\left( z\right) =\sqrt{\mathcal{X}^{2}}\Psi
\left( 1/2,-\left\vert m\right\vert +1/2;\mathcal{X}^{2}\right) $ for
several choices of $\mathcal{B}$\ and $z$. Because the CHF is a
monotonically increasing function of $\mathcal{B}$, the~inequality $%
R^{\left( \left\vert m\right\vert \right) }\left( z\right) \geq \left.
R^{\left( \left\vert m\right\vert \right) }\left( z\right) \right\vert _{%
\mathcal{B}=10^{5}}$ holds for each fixed choice of $z$. Thus, selecting $%
z=1.5\lambdabar _{\mathrm{C}}$ for example, we~calculate the ratio $%
R^{\left( 0\right) }\left( 1.5\lambdabar _{\mathrm{C}}\right) $ and conclude
that the effective potential (\ref{ad1.15}) is Coulombian (\ref{ad1.15c})
with an accuracy no less than $93\%$. Moreover, the~accuracy decreases as $%
\left\vert m\right\vert $ increases\footnote{%
This~is not unexpected, because the shallower the effective potential, the
more it \textquotedblleft deviates\textquotedblright\ from the Coulombian
pattern at small distances, as~can be seen in Figure~\ref{F1}.}, for
instance $R^{\left( 1\right) }\left( 1.5\lambdabar _{\mathrm{C}}\right) >0.87
$, $R^{\left( 2\right) }\left( 1.5\lambdabar _{\mathrm{C}}\right) >0.83$, $%
R^{\left( 3\right) }\left( 1.5\lambdabar _{\mathrm{C}}\right) >0.78$. At $%
z=2\lambdabar _{\mathrm{C}}$, we~obtain $R^{\left( 0\right) }\left(
2\lambdabar _{\mathrm{C}}\right) >0.96$, $R^{\left( 1\right) }\left(
2\lambdabar _{\mathrm{C}}\right) >0.92$, $R^{\left( 2\right) }\left(
2\lambdabar _{\mathrm{C}}\right) >0.89$, and~$R^{\left( 3\right) }\left(
2\lambdabar _{\mathrm{C}}\right) >0.86$. Finally, it remains to justify the
approximation for the logarithmic derivative of the wave function at small
distances, given by Equation~(\ref{ad1.18}). This~expression relies on the
assumption that $\varepsilon _{\parallel }\rho ^{2}\left( \varepsilon
_{\perp }a_{\mathrm{B}}^{2}\xi ^{2}\right) ^{-1}=\varepsilon _{\parallel
}\rho ^{2}\left( \varepsilon _{\perp }z^{2}\right) ^{-1}\ll 1$\ is satisfied
for distances within the short-range interval. To check it, we~replace $\rho 
$ by the magnetic length $a_{\mathrm{H}}$, $\rho =a_{\mathrm{H}}$,\ and
substitute $z$ by its highest value within $z\in \left[ 0,\ell \right] $,
i.e., $z=\ell $. For $\ell =1.5\lambdabar _{\mathrm{C}}$, the~above parameter
varies approximately from $8.4\times 10^{-2}$ to $3.5\times 10^{-4}$, for
magnetic fields within the range (\ref{range}). As~a result, there are no
restrictions on the applicability of the approximation (\ref{ad1.20}) in the
estimate (\ref{ad1.18}).

\begin{figure}[h!]\centering
\includegraphics[scale=0.4]{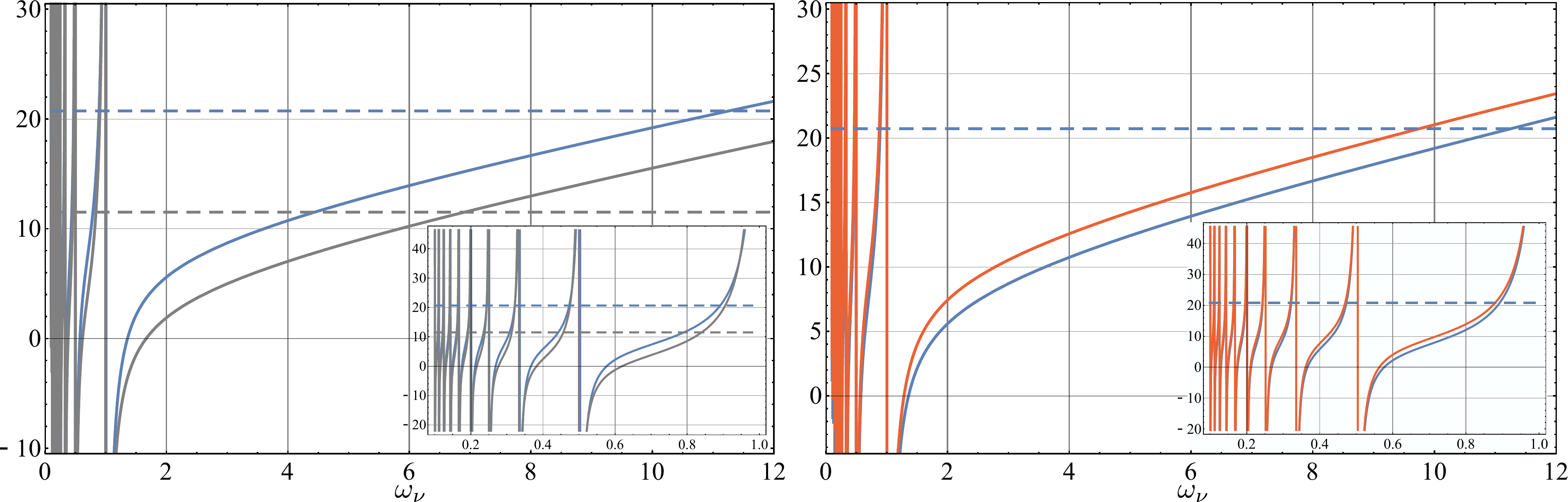}
\caption{(color online) Energy levels of the Hydrogen
atom in strong magnetic fields. The~left panel illustrates only the lowest
levels ($m=0$) for $\mathcal{B}=10^{5}$ $\left( b\approx 5.3\right) $ (solid gray lines) and for $\mathcal{B}=10^{9}$ $\left( b\approx 5.3\times
10^{4}\right) $ (solid blue lines). In~the right panel, all lines were
computed for $\mathcal{B}=10^{9}$, where the blue ones correspond to $m=0$
while the red ones to $m=3$. In~both panels, the~horizontal dashed lines
refer to $\ln 10^{5}\approx 11.5$ (in gray) and $\ln 10^{9}\approx 20.7$ (in blue).}
\label{F2}
\end{figure}

Based on these results, we~conclude that the modified KP equation (\ref{ad2}%
) meets the requirements needed to estimate energies for lower even levels.
To estimate the energies numerically, we~conveniently represent the
transcendental Equation (\ref{ad2}) graphically for some fixed values of the
external field and of the quantum number $m$. This~is illustrated in Figure %
\ref{F2}, where solid lines represent the right-hand side of Equation~(\ref{ad2}),
while the horizontal dashed lines represent the left-hand side, i.e., $\ln \mathcal{B}$%
. For a given set of values to $\mathcal{B}$ and $m$, one obtains an
infinite number of roots of Equation~(\ref{ad2}) for $\omega _{\nu }<1$, but
single roots for $\omega _{\nu }>1$, corresponding to the deep energy
levels; the~deepest one owing to $m=0$, as~can be seen on the right panel
of Figure~\ref{F2}.

\begin{figure}[h]
\centering
\includegraphics[scale=0.6]{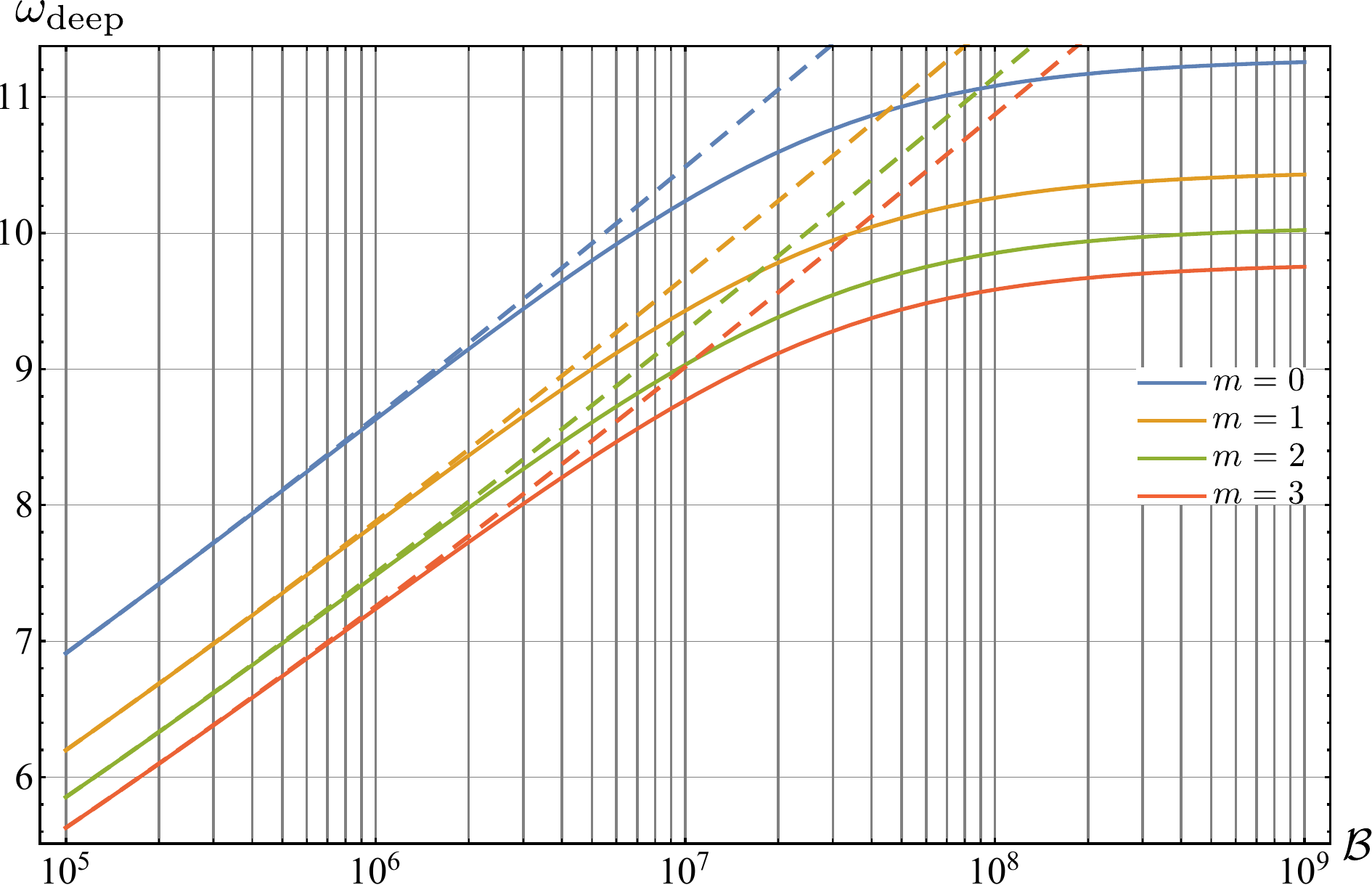}
\caption{(color online) Deepest energy levels (in Rydberg units) of even
states of the Hydrogen atom in the local approximation (solid lines) and
neglecting VP (dashed lines).}
\label{F3}
\end{figure}

In~Figure~\ref{F3}, we~illustrate numerical results for the deeper lower energy
levels as functions of the magnetic background $\mathcal{B}$ according to
Equation~(\ref{ad2}) (solid lines). In~this same figure, we~also include
numerical results using the Karnakov--Popov equation without the vacuum
polarization (dashed~lines).

In~contrast to cases where VP is neglected, we~observe from
Figure~\ref{F3} that the energy levels \textquotedblleft
saturate\textquotedblright\ to specific values as the magnetic field grows.
The~saturation occurring at sufficiently strong magnetic fields can be
computed directly from Equation~(\ref{ad2}) after the asymptotic expressions for
the longitudinal dielectric permittivity (\ref{As}) are substituted into (%
\ref{ad2})%
\begin{equation}
\frac{\omega _{\mathrm{sat}}}{Z}+2\ln \omega _{\mathrm{sat}}+2\psi \left( 1-%
\frac{Z}{\omega _{\mathrm{sat}}}\right) \approx \ln \left( \frac{\mathcal{B}%
}{1+\alpha ^{3}\mathcal{B}/3\pi }\right) -4\gamma -\ln 2-\psi \left(
\left\vert m\right\vert +1\right) \,.  \label{n4}
\end{equation}%

The~above expression surprisingly coincides with the modified KP equation
obtained beyond the local approximation, reported previously by Machet and
Vysotsky in~\cite{MacVys11}. This~explains why saturation values obtained
within the local approximation%
\begin{equation}
\omega _{\mathrm{sat},m=0}\approx 11.213\,,\ \ \omega _{\mathrm{sat}%
,m=1}\approx 10.393\,,\ \ \omega _{\mathrm{sat},m=2}\approx 9.987\,,\ \
\omega _{\mathrm{sat},m=3}\approx 9.719\,,  \label{n5}
\end{equation}%
coincide with results obtained in that work. The~corresponding saturation
energies $\mathcal{E}_{\mathrm{sat}}=-\omega _{\mathrm{sat}}^{2}\times 
\mathrm{Ry}$ read $\mathcal{E}_{\mathrm{sat},m=0}\approx -1.71\mathrm{KeV}$, 
$\mathcal{E}_{\mathrm{sat},m=1}\approx -1.47\mathrm{KeV}$, $\mathcal{E}_{%
\mathrm{sat},m=2}\approx -1.36\mathrm{KeV}$, and~$\mathcal{E}_{\mathrm{sat}%
,m=3}\approx -1.29\mathrm{KeV}$. An explanation for the similarity between our results and those beyond the local approximation can be provided by analyzing the behavior of the logarithmic derivative of the wave functions at short distances. In~fact, substituting the Yukawa-like potential\footnote{%
to which the Coulomb potential is modified by the VP beyond
the local approximation~\cite{ShaUso07,ShaUso071,sadooghi} in the form it
acquires with the help of the interpolation of~\cite{MacVys11}.} in the
definition of the effective potential~(\ref{ad1.4c}) and performing the same
approximations as the ones described in the previous section, the~authors of 
\cite{MacVys11} found the following expression for the logarithmic derivative%
\begin{equation}
\frac{\chi _{0m}^{\nu \prime }\left( \xi \right) }{\chi _{0m}^{\nu }\left(
\xi \right) }\approx -Z\left[ \ln \left( \frac{2\mathcal{B}}{1+\alpha ^{3}%
\mathcal{B}/3\pi }\xi ^{2}\right) -\psi \left( \left\vert m\right\vert
+1\right) \right] \,,  \label{n6}
\end{equation}%
which coincides with our result Equation~(\ref{ad1.18}) in the limit of strong
magnetic field, thanks to the asymptotic properties of the dielectric
permittivities in this case (\ref{As}). According to these results, we
conclude that the saturation of energy levels of the hydrogen atom---known
to exist when the atom is exposed to a sufficiently strong magnetic field 
\cite{ShaUso07,ShaUso071,MacVys11}---can also be obtained within the local
approximation.

\section{Conclusions}

The~linear growth of the polarization tensor with the external magnetic
field~\cite{skobelev,skobelev2,melrose,Shabad88,heyl} results
in strong screening of the static field of a point-like electric charge \cite%
{ShaUso07,ShaUso071}. It affects the one-dimensional Schr\"{o}dinger equation responsible
for the longitudinal motion of an electron in a hydrogen atom placed in a
very strong magnetic field. As~a result, the~Coulomb singularity $\sim
r^{-1}$ close to the charge is deformed. When the polarization tensor is
calculated~\cite{BatSha71,BatSha712,BatSha713,BatSha714,BatSha715} as one loop of electron propagators taken for
Green functions of the Dirac equation in a magnetic field, the~above behavior
is replaced by a Yukawa-like one~\cite{ShaUso07,ShaUso071,sadooghi}, which
turns into $\delta (x)$ as the magnetic field tends to infinity. Contrary to
the singularity $r^{-1}$ that leads to an infinite sinking of the ground
state energy in that limit, known in quantum mechanics~\cite{EllLou60,EllLou6066}, the~delta function appearing due to the radiative corrections leaves the
ground state energy finite~\cite{ShaUso07,ShaUso071}. This~is called \textquotedblleft
saturation\textquotedblright\ or \textquotedblleft
freezing\textquotedblright\ of the level. Its specific value was estimated
differently in~\cite{ShaUso07,ShaUso071} and in~\cite{MacVys11,PopKar12,PopKar14}. The~latter authors used the \textquotedblleft KP
method\textquotedblright\ developed previously in~\cite{PopKar03} to find the hydrogen spectrum in a strong magnetic field, which improves the
calculational results of QM relying on pure adiabatic approximation, when
only the diagonal part of the matrix effective potential is used. In~applying
this method to QED, a simplified interpolation of the polarization tensor
was exploited in~\cite{MacVys11} that allowed to perform analytical
calculations to determine large-field asymptotic value of the ground state
(as well as excited states), different from the value found in~\cite{ShaUso07,ShaUso071}
by a different method.

In~the present paper, we~establish that another approximation for the
polarization operator---physically meaningful, contrary to the interpolation
of~\cite{MacVys11}---can be used analytically following the same KP method,
that also leads to the saturation.

This~is just the Euler--Heisenberg~\cite{HeiEul36} local action approximation
LCFA, from which the polarization tensor covering the screening of
slowly-varying fields is obtained by differentiations over its field
arguments. The~asymptotic long-range behavior of the potential (\ref{n1}) of
a point-like charge and the corresponding transverse and longitudinal
dielectric permittivities $\varepsilon _{\perp }$,\ $\varepsilon _{\parallel
}$ are determined by eigenvalues of this polarization tensor. Their leading
large-field asymptotes are given in (\ref{As}). Similarly to the fact known
beyond LCFA, the~growth, linear with the magnetic field, is peculiar to the
polarization tensor within this approximation and shows itself as the growth
of the longitudinal dielectric permittivity $\varepsilon _{\parallel }$ in (%
\ref{As}) responsible for a growing screening of the charge. The~mechanism
of how this large screening results in finiteness of the ground-state energy
is, however, different within LCFA. It reduces to the change of the minimum
size of the wave function, the~magnetic length $a_{\mathrm{H}}=%
\sqrt{\hslash c/eB}$---that in QM tends to zero when $B$ increases and
thereby supplies the infinity to the potential of the point charge---by the
value $\sqrt{\varepsilon _{\parallel }}a_{\mathrm{H}}$ that has a finite
limit $\sqrt{\alpha /3\pi }\lambdabar _{\text{\textrm{C}}}$. This~finite
limit makes a sort of elementary length, such that the electron cannot
approach the nucleus closer than it. Therefore, the~Coulomb singularity is
regularized (see (\ref{Pot}) or (\ref{ad1.15})), and~the LCFA is better
justified. Our~final result for the ground state turns out to be the same as
beyond this approximation.

Last, but not least, one ought to say that this nontrivial congruence---between results obtained within and beyond the local approximation---can be observed on other physical problems. For example, recently, the~running of the fine-structure constant and the electron dynamical mass in a constant magnetic field has been considered beyond the local approximation in~\cite{FerSan19}. The~final reported results for the fine-structure constant for magnetic fields within the range (\ref{range}) can be equally obtained within the local approximation because the eigenvalue of the polarization tensor in the infrared approximation $\varkappa_{2}=\varepsilon_{\perp}\mathbf{k}_{\perp}^2+\varepsilon_{\parallel}k_{\parallel}^{2}$~\cite{ShaUso11} behaves asymptotically according to Equation~(\ref{As}), capturing thereby the same results of Ref.~\cite{FerSan19}. As~a result, the~electron dynamical mass calculated within or beyond the local approximation remains the same.

\section*{Acknowledgments}
This research was funded by the Russian Science Foundation, grant number 19-12-00042.

% The~following MDPI journals use author-date citation: Arts, Econometrics, Economies, Genealogy, Humanities, IJFS, JRFM, Laws, Religions, Risks, Social Sciences. For those journals, please follow the formatting guidelines on http://www.mdpi.com/authors/references
% To cite two works by the same author: \citeauthor{ref-journal-1a} (\citeyear{ref-journal-1a}, \citeyear{ref-journal-1b}). This~produces: Whittaker (1967, 1975)
% To cite two works by the same author with specific pages: \citeauthor{ref-journal-3a} (\citeyear{ref-journal-3a}, p. 328; \citeyear{ref-journal-3b}, p.475). This~produces: Wong (1999, p. 328; 2000, p. 475)

%=====================================
% References, variant B: external bibliography
%=====================================
%\externalbibliography{yes}
%\bibliography{your_external_BibTeX_file}

%%%%%%%%%%%%%%%%%%%%%%%%%%%%%%%%%%%%%%%%%%
%% optional

%% for journal Sci
%\reviewreports{\\
%Reviewer 1 comments and authors’ response\\
%Reviewer 2 comments and authors’ response\\
%Reviewer 3 comments and authors’ response
%}

%%%%%%%%%%%%%%%%%%%%%%%%%%%%%%%%%%%%%%%%%%
\end{document}